# Mergers and Galaxy Evolution

*R. G. Carlberg*

Department of Astronomy, University of Toronto, Toronto M5S 1A7, Canada

## 1 Summary

Galaxy merging is the late time manifestation of the galaxy formation process and likely significantly effects $z < 1$ galaxies. A "maximum reasonable rate" model for merging finds a $\sim 2$ mag K band increase in the luminosities of dwarf galaxies so that they contribute significantly to the faint counts, with spirals and ellipticals being far less affected. The median $K$ and $I$ redshifts stabilize (and even decrease slightly) at $z \simeq 0.6$ beyond I=21 or K=19. The B redshifts continue to rise (although strongly dependent on the UV spectral evolution). Such rapid merging predicts that at $z = 1$ the characteristic galaxy mass is reduced to $\sim 30\%$ of the $z = 0$ value. To rule out this model requires good sampling beyond $z = 1$.

A theoretical complication for even a minimal merger rate, which reduces $z = 1$ masses to 2/3 of current epoch values, is that infall of a single satellite having 10% of a disk's mass may destroy thin disks. Using completely self-consistent n-body simulations, we show that the primary response of a disk to "cosmological" satellites up to 20% of the disk mass is to *tilt* the disk with a temporary warping.

## 2 Merger Rates

The inevitability and significance of merging was highlighted in an influential article [T] which emphasized merging as a source of transformation within the Hubble sequence from disk to spheroidal galaxies. Most aspects of Toomre's argument remain valid today. First, long tidal tails drawn from a *stellar* disk are reliable indicators that the galaxies are on relative orbits sufficiently energetically bound that dynamical theory (and a huge amount of simulation data) indicates the two galaxies will merge within 1-2 rotation periods, typically 0.5 Gyr. Second, the merger rate is expected to rise into the past, Toomre giving an energy distribution argument for a rate increase of $(1+z)^{5/2}$. Third, a lower limit on the present day merger rate is 11 completely "doomed" pairs in approximately 4000 galaxies, which translates to a rate of approximately $0.005$ $Gyr^{-1}$. Integrated to $z = 1$ this reduces the mass of an average galaxy about 10% (for $t_0 \simeq 13$ Gyr). Hence, basic issues are: 1) how to recognize a merger, 2) the current epoch merger rate, and 3) the redshift dependence of the merger rate.

A somewhat separate line of investigation is to predict the outcome of realistic merger events. A "major merger" (nearly equal mass galaxies) inevitably creates a system with essentially no disk and more stars on the orbits characteristic of a spheroidal galaxy [BH92]. The result of a "minor merger", involving a disk system such as the upcoming LMC-Galaxy merger, is considerably less certain, and is discussed below. Star formation and ISM alterations are less certain in general, but a general fate for disk gas is to be reverse torqued by a merger induced bar, and very quickly moved into the central region [BH91].

## 2.1 Empirical Merger Rates

At low redshift, estimates of the merger rate fall into two categories. Morphological methods use images of galaxies to determine which systems soon will soon merge and which ones are likely merger remnants (which gives the time integral of the merger rate). These methods are strongly dependent on a "training set" of merging galaxy images, recently considered for HST images [M95]. Estimates using this technique use the Arp Atlas [T] to give $0.005$ $Gyr^{-1}$ and the Arp & Madore Atlas [CC] with a looser requirement for "tidal features" gives $0.02$ $Gyr^{-1}$. The difficulty with morphological classification for mergers is that the best indicators depend on tails and shells, both of which are generally very low surface brightness features which require imaging data of very high quality. Furthermore, there will almost certainly be phases of the merger which are not morphologically recognized as a merger-in-progress [M95].

An alternate empirical estimate is to use the basic dynamical result that all pairs of galaxies separated by a galaxy diameter or less and having a relative velocity less than the escape velocity at the edge of the disk will merge in 1-2 orbital times, under the important assumption that external tides can be neglected. A useful, but somewhat arbitrary, definition of the maximum separation to constitute a close pair is $20h^{-1}$ kpc. At this distance visible disks will interact significantly, and, external tides should be insignificant. Using this definition, the UGC catalogue of nearby galaxies has 2.3% of its galaxies in low velocity, close pairs, implying a merger rate of $0.046$ $Gyr^{-1}$ [CPI]. At $z \simeq 0.4$ the fraction of close pairs (with little information about relative velocities) is 8.5 to 10% [CPI, YE95], implying a strong rise in the merger rate, $\sim (1+z)^4$. These studies find no evidence for the expected blueing of the colours of close pairs [LT], although this could be a consequence of dilution with projected galaxies, and the absence of multi-band data.

## 2.2 Theoretical Merging Rates

A semi-empirical approach to estimate the merger is to calculate the rate of dynamical friction inspiral. The density of neighbours comes from the galaxy-galaxy correlation function. Only those pairs which are moving at relative velocities such that they are on nearly parabolic orbits merge. That is, rich galaxy clusters have few internal mergers, in spite of the high galaxy density. The basic approach is to compute the rate of inspiral of companion galaxies taking the circular velocity from a Fisher-Tully relation, the mass of companions from the galaxy luminosity function, the density of companions from the galaxy correlation function, and an assumption about the initial orbital ellipticity. For reasonable choices of these parameters, these calculations find that to a time of 5 Gyr a typical galaxy will have merged with approximately $20 \pm 10\%$ of its mass [BT88, TO, H95], implying a merger rate of 0.04 Gyr$^{-1}$. This rate is about twice what one infers from morphological indicators, but, not all merging galaxies are expected to have clearly recognizable merger characteristics. Extrapolating (at a uniform rate in time) to redshift 1, this implies that an average galaxy will have a mass that is 67% of the present day values.

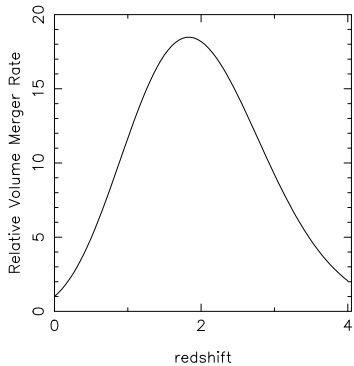

Figure 1: The volume merger rate as a function of redshift, relative the redshift zero rate. The location of the peak is set by the fluctuation amplitude on galaxy scales. The peak can be interpreted as the time of assembly of most galaxies. The speed of the rise at low redshift depends on $\Omega$ and the fraction of pairs moving slowly enough to be gravitationally captured.

A theoretical merger rate was derived in [C90], which a superior later calculation [LC] finds to be an overestimate for dark matter halos of galaxies. The [LC] calculation finds that to redshift 1 the average galaxy halo mass will be reduced to about 35% of its present day value [NFW], which serves as a useful upper limit (and possibly the correct value) to the merger rate of galaxies, since galaxies can merge no faster than their dark halos. The situation for the visible components of galaxies, which are likely not well described by the Press-Schechter theory [PS], is far less clear. For a practical calculation of the effect of merging on galaxies, the rate will be normalized using the empirical data, extrapolated using my functional form, which to a redshift of 0.5 is approximately equal to $(1 + z)^4$, flattening off at higher redshifts as illustrated in Figure 1.

## 3 Merging and Faint Galaxies

The impact of merging on galaxy properties at earlier epochs is potentially large, however, at the moment there is relatively little evidence for any change in the underlying bulk stellar populations of galaxies [CFRS]. The morphological dependence of the merger models of [CCh, C92] are displayed below as an example of a "maximum merger model". As others [CSH, BEG] have found, a significant change in the galaxy population, as would be expected from rapid merging is consistent with quite a lot of the data. The idea here is to push the model far enough that a conflict with observational data may emerge. The model has a luminosity dependent density evolution, arising from the increase in gas content to later Hubble types. Mergers are assumed to induce a starburst in much of this gas, resulting in evolution of the luminosity function along the lines suggested on other grounds [BES, CSH, BEG]. The main parameter of interest here is the redshift zero merger rate, which is set at $0.04\,\mathrm{Gyr}^{-1}$, as indicated by the pair counts and dynamical friction infall calculations discussed above. The counts as a function of morphological type are shown in Figure 2. The rapid rise in the dwarf population is similar to the HST data [GESG]. Of particular note is the plateau in the median redshift for red selected galaxies. The $n(z)$ does remain broad, with extending to $z = 2$ or so with reasonable numbers.

**Table 1.** Maximal Merging Median redshifts

| B $z_{1/2}$ | V $z_{1/2}$ | I $z_{1/2}$ | K $z_{1/2}$ |
|---|---|---|---|
| 20 0.18 | 20 0.21 | 17 0.26 | 15 0.20 |
| 21 0.25 | 21 0.26 | 18 0.36 | 16 0.31 |
| 22 0.34 | 22 0.31 | 19 0.47 | 17 0.45 |
| 23 0.45 | 23 0.39 | 20 0.57 | 18 0.59 |
| 24 0.55 | 24 0.50 | 21 0.62 | 19 0.65 |
| 25 0.64 | 25 0.61 | 22 0.61 | 20 0.60 |
| 26 0.71 | 26 0.71 | 23 0.59 | 21 0.52 |
| 27 0.77 | 27 0.72 | 24 0.58 | 22 0.50 |
| 28 0.81 | 28 0.71 | 25 0.57 | 23 0.51 |
| 29 0.83 | 29 0.75 | 26 0.57 | 24 0.54 |
| 30 0.84 | 30 0.78 | 27 0.58 | 25 0.57 |

### 3.1 Recapitulation

This "maximal merging" calculation implies such drastic evolution beyond redshift one (galaxies have a characteristic masses only 30% of the current epoch values, and the median redshift are very low) that compared to a no-merger model the differences are easily detectable in surveys beyond $K \sim 20$ (bands of $I$ and bluer having reduced sensitivity since the 4000Å break crosses through at $z \simeq 1$). Lower redshift data is less clearly interpreted, since mass decreases are partially offset with luminosity increases. If (or when) the predictions are found to be in conflict with observations, the main implication is that the extrapolation of the merging rate with redshift is too large. The simplest and most natural way to make a substantial change in the merger rate is to put $\Omega \simeq 0.2$ [C90], although this requires a wholesale re-examination of many aspects of observational cosmology for consistency. It should be noted that even a constant merger rate predicts that galaxy masses at $z = 1$ will be reduced to 67% of current epoch values.

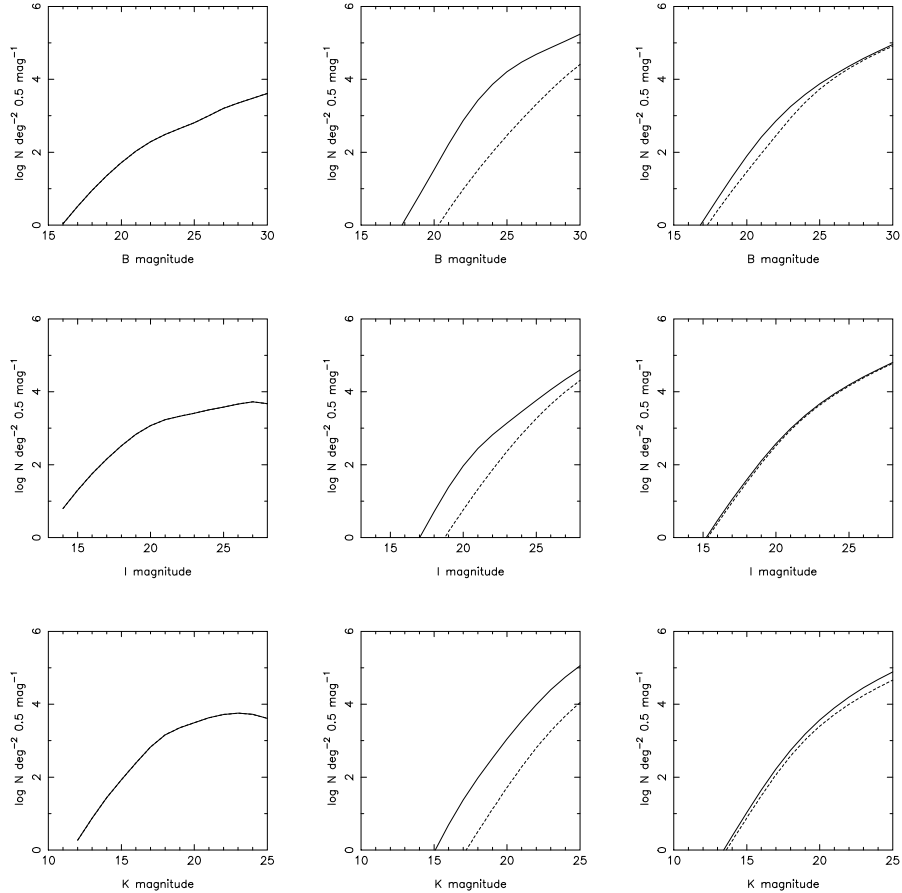

Figure 2: The dependence of faint galaxy counts on morphology. The first column is E/S0, the second Irr, the third S. The dashed line is evolution with merging but no luminosity changes, the solid line shows the luminosity enhancements due to starbursts. E/S0 types are assumed to have no gas. The numbers Irr types become comparable to S types around $I = 22$.

## 4 Merging and Thin Disks (with Siqin Huang)

Galactic disks are remarkably thin, typically having a ratio of horizontal to vertical scale lengths of about 10 to 1, which implies that the ratio of kinetic energy in horizontal to vertical motions is about 100 to 1. Hence globally small energy additions could give dramatic vertical structure changes. Several calculations have found that a satellite having 10% of the disk mass spiraling into the disk [QG, QHF, TO] will unacceptably thicken the disk. As noted above, straightforward dynamical calculations find that a typical disk must absorb about $2 \pm 1$ such satellites. There is no doubt that if a 10% satellite (possibly even a 1% satellite) enters a disk, the disk will be unacceptably thickened.

There are several reasons to believe that thin disks may be quite robust, as long as the satellite substantially dissolves prior to entry into the disk. The strong self-gravity of a disk causes all of its dynamical frequencies to be much higher than the perturbing frequencies of the satellite. For instance, in the solar neighbourhood of our galaxy the ratio of the vertical oscillation frequency to the orbital frequency is a factor of 5, so an exterior satellite would be seen as an adiabatic potential variation. Nevertheless the satellite does exert a torque on the disk stars. The issues are whether a satellite is greatly tidally stripped before it reaches the disk, and, whether the disks stars respond incoherently (leading to disk thickening), or, coherently (leading to some combination of warp/tilt/precession). In particular, it has been suggested [B90] that warps indicate the recent addition of misaligned angular momentum.

Our approach to the problem is in the spirit of a reasonable counterexample, the calculation being done with a completely self-consistent n-body simulation. This is a challenging problem for n-bodies, mainly because large numbers of particles are needed in order to keep unwanted two-body heating at a level below the dynamics of interest, and, the simulation must be run for many disk rotations (because satellite infall is slow compared to disk dynamical times). The simulations below have been done with particle numbers ranging from 60,000 to 200,000 to check for two-body effects, shorter time steps, and an alternate time stepping scheme. No substantial differences in the conclusions result. The system consists of a disk, a halo, and a satellite, all composed of active particles. The "live" halo is particularly important, since in its absence the transfer of angular momentum from the disk to the satellite drives the satellite away (unless the satellite is started very close to the disk [QG]). The heating of the disk by the halo is significant, but comparable (or less than) the observed rate of increase in the velocities of disk stars. The satellite structure is derived from "cosmological" initial conditions. That is, it is assumed that the density profile of the galaxy's halo and the satellite are self-similar, with the velocity dispersions scaling as $\sigma \propto M^{1/3}$. This is a crucial assumption, which others have generally not made. It implies, whatever the mass profile, that once the satellite has spiraled through 90% of the galaxy's dark halo, say from 100kpc down to 10kpc, then the satellite will have had about 90% of its mass tidally removed, greatly reducing its impact on the disk.

We find that the primary response of a thin disk to "cosmological" low mass satellite infall (up to 20% of the disk mass) is to develop a temporary warp, tilt over (with little precession [NT]) all with negligible thickening. The outcome for a particular satellite will depend a great deal on the mass profile of the satellite itself. Only the rare, dense dwarfs, such as M33, will be able to cause a lot of disk disruption. Noting these caveats, our conclusion is that the predicted infall rate of typical low density dwarfs can be accommodated without unreasonable disk thickening.

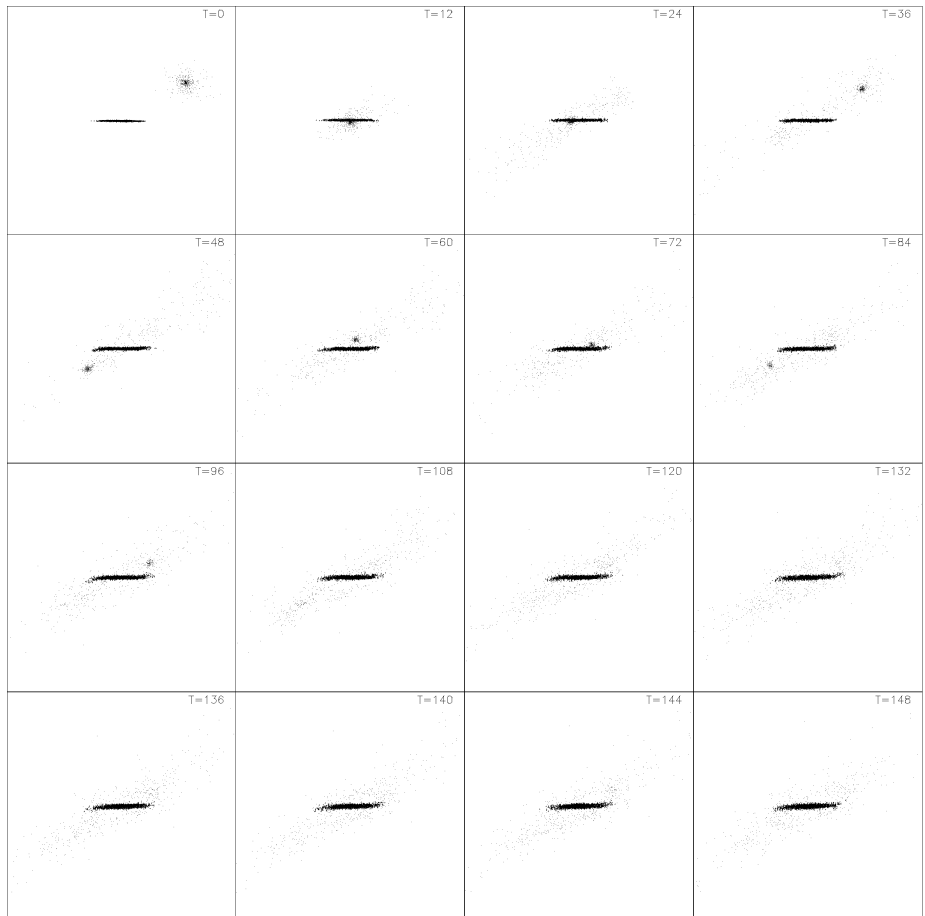

Figure 3: The infall of a 10% satellite onto a disk. The halo particles are not plotted. The primary response of the disk is a tilt of about 4° with essentially no precession. A transient warp is visible. A counter-rotating satellite tilts the disk about half as much in the other direction.

## 5  Conclusions

The role of merging in the evolution of galaxies must be significant, but the details remain unclear. This situation will change dramatically over the next few years. First, deeper redshift surveys will soon constrain any dramatic evolution of the galactic mass function. Second, detailed studies of individual galaxies at moderate redshift will measure their scale lengths and kinematic properties to directly examine masses to check whether they are dramatically lower. Third, simulations now have the dynamical resolution to answer questions involving the merging of galaxies in a cosmologically realistic setting. In particular, simulations can usefully guide morphological studies of faint galaxies and constrain the redshift-merger rate relation. It should be emphasized that if the time to $z = 1$ is $\simeq 8.5$ Gyr, and if the merger rate remains constant at $0.04$ Gyr$^{-1}$, then galaxy masses will at most be 67% of current epoch values. Studies of the evolution of clustering of galaxies will be key components of constraining the merger-redshift relation. Unless the understanding of dark halo densities and dynamics is severely flawed mergers *must* be significant at higher redshifts.

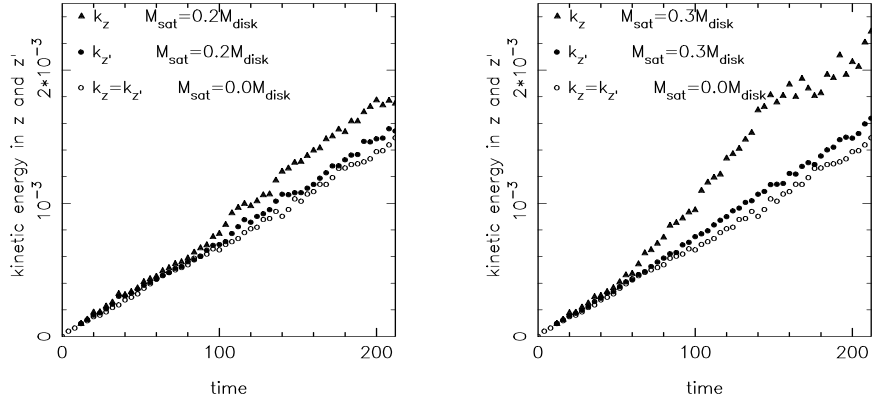

Figure 4: Disk kinetic energy in vertical motions versus time for 20% and 30% satellite inspiral with a no-satellite control model. The vertical motions are calculated in the initial frame and the current principle axis frame ($z'$). At the initial disk half mass radius the circular velocity is nearly 1 unit and the rotation period about 4 time units. The 20% satellite barely heats the disk. The 30% satellite causes about a 20% thickening inside the half mass radius, and a doubling to tripling beyond that radius. The energy is mainly absorbed as a coherent tilting of the disk.

**Acknowledgements:** The disk galaxy simulations are are a substantial part of Siqin Huang's Ph. D. thesis work. This research is supported by NSERC of Canada.